\documentclass[aps,prl,showpacs,floatfix,superscriptaddress,nofootinbib,10pt,twocolumn]{revtex4-2}
\usepackage{color,amsmath,amssymb,graphicx,latexsym,subfigure}
\usepackage{threeparttable,multirow,txfonts,makecell,tabularx}

\begin{document}
	
	\title{Gravitational laser: the stimulated radiation of gravitational waves from the clouds of ultralight bosons}
	
	\author{Jing Liu}
	\email{liujing@ucas.ac.cn}
	
	\affiliation{International Centre for Theoretical Physics Asia-Pacific, University of Chinese Academy of Sciences, 100190 Beijing, China}
	
	\affiliation{Taiji Laboratory for Gravitational Wave Universe~(Beijing/Hangzhou), University of Chinese Academy of Sciences, 100049 Beijing, China}
	
	\begin{abstract}
		Stimulated radiation and gravitational waves (GWs) are two of the most important predictions made by Albert Einstein. In this work, we demonstrate that stimulated GW radiation can occur within gravitational atoms, which consist of Kerr black holes and the surrounding boson clouds formed through superradiance. The presence of GWs induces mixing between different states of the gravitational atoms, leading to resonant transitions between two states when the GW wavenumber closely matches the energy difference. Consequently, the energy and angular momentum released from these transitions lead to the amplification of GWs, resulting in an exponential increase in the transition rate. Remarkably, the transitions complete within a much shorter time compared to the lifetime of the cloud. These stimulated transitions give rise to a novel GW signal that is strong and directed, distinguished from the previously predicted continuous GWs originating from clouds of ultralight bosons.
	\end{abstract}
	
	\maketitle
	\emph{Introduction}. 
	Ultralight fields that weakly interact with the Standard Model of particle physics, such as axions~\cite{Weinberg:1977ma,Arvanitaki:2009fg,Irastorza:2018dyq} and dark photons~\cite{Holdom:1985ag,Goodsell:2009xc,Cicoli:2011yh,An:2022hhb}, are promising candidates for dark matter~\cite{Hu:2000ke,Bertone:2004pz,Essig:2013lka}. When the Compton wavelengths of these ultralight fields are comparable to the size of the event horizon of a black hole~(BH), they can extract angular momentum and energy from the BH through superradiance~\cite{Zouros:1979iw,Detweiler:1980uk,Dolan:2007mj,Arvanitaki:2010sy,Brito:2015oca,Endlich:2016jgc,Baumann:2019eav,Bao:2022hew}, leading to the formation of dense clouds. This system, known as a ``gravitational atom", is similar to a hydrogen atom.
	The annihilation of these ultralight fields and transitions between different levels of gravitational atoms produce nearly-monochromatic gravitational waves (GWs)~\cite{Arvanitaki:2010sy,Yoshino:2013ofa}. With recent advancements in GW observation techniques, an intriguing opportunity arises to explore and constrain the dark sector of the Universe by observing GWs emitted from these boson clouds~\cite{Arvanitaki:2014wva,Brito:2017zvb,Brito:2017wnc,Hannuksela:2018izj,Isi:2018pzk,Palomba:2019vxe,Zhu:2020tht,Yuan:2022bem}. Analogous to stimulated radiation in electromagnetic interactions, an interesting question emerges: can stimulated radiation occur within gravitational atoms? 
	In previous works, the authors find the additional gravitational potential from a companion of the BH can trigger resonant transition from level mixing, and the angular momentum of the cloud is released into the orbital angular momentum of the companion~\cite{Baumann:2018vus,Zhang:2018kib,Zhang:2019eid,Berti:2019wnn,Takahashi:2021eso,Tong:2022bbl,Cao:2023fyv,Fan:2023jjj,Guo:2024iye}.
	In this work, we find that external GWs can also induce level mixing. Notably, GWs with wavenumbers close to the energy level difference lead to resonant transitions, which in turn amplify the strength of GWs, thereby accelerating the transition rate. Consequently, ultralight bosons transition into unstable states within a short period compared to the cloud's lifetime, releasing most of the energy between the levels into GWs. This physical phenomenon resembles superfluorescence~\cite{PhysRev.93.99}, where the transition of minority particles amplifies the transition rate of the other particles.
	As the strength of GWs diminishes rapidly after the transitions, the particles in the lower unstable states are eventually absorbed by the BH instead of oscillating between the two levels. This mechanism offers a novel prediction of strong and directed GWs, distinct from the well-known continuous GWs emitted by boson clouds. Such GW signals are expected to be detectable by various GW detectors~\cite{TheLIGOScientific:2014jea,Somiya:2011np,Audley:2017drz,Guo:2018npi,Carilli:2004nx}, providing a new method to investigate the dark sectors of new physics.
	For convenience, we choose $\hbar=c=G=1$ throughout this $letter$.

	\emph{Superadiance and boson clouds}.
	Superradiant instability leads to the formation of boson clouds around rotating BHs. Consider a system consisting of a Kerr BH with mass $M$ and an ultralight scalar field $\Psi(t, \mathbf{r})$ with mass $\mu$. The system is governed by the gravitational interaction.
	
	In the non-relativistic limit, the Klein-Gordon equation of $\Psi(t, \mathbf{r})$ in a Kerr spacetime can be transformed into a form similar to the Schrödinger equation of the hydrogen atom. The parameter $\alpha = M\mu$ represents the fine-structure constant of the gravitational atom. See the appendix for more details.
	The eigenfrequencies are given by
	\begin{equation}\label{eq:eigenf}
		\omega_{n \ell m} \approx E_{nlm}+i\Gamma_{nlm}\,.
	\end{equation}
	where $n$, $l$ and $m$ are the principal, azimuthal and magnetic quantum numbers, respectively. The imaginary term $\Gamma_{nlm}\propto\left(m \Omega_H-\omega_{n l m}\right) \alpha^{4l+5}$~\cite{Detweiler:1980uk} arise from the boundary conditions at the BH horizons where $\Omega_H$ is the angular velocity of the outer horizon. This is an important difference from the hydrogen atom. Eq.~\eqref{eq:eigenf} implies the occupation number in the states with $0<\omega_{nlm}<m\Omega_{H}$ grows exponentially or otherwise absorbed into the BH. The superradiant rate of the states with large $l$ is strongly suppressed by the dependence of $\alpha^{4l+5}$. As the angular momentum of the BH is transferred into the boson cloud, $\Omega_{H}$ gradually decreases and the superradiance terminates when $m\Omega_{H}\leqslant\omega_{nlm}$ becomes smaller than $\omega_{nlm}/m$. 
	
	The annihilation of the ultralight fields can generate continuous monochromatic GWs with a wavenumber $k=2\mu$~\cite{Yoshino:2013ofa}, which is the primary GW production channel for low $l$. The timescale of this GW production is in general much longer than that of superradiant instabilities. However, as we will see, the existence of external GWs can result in resonant transitions and the depletion of the cloud in a much shorter period.

	
	\emph{The stimulated radiation of GWs}.
	The existence of external GWs introduces an interaction term to the Hamiltonian, $H_{I}$. To the linear order of tensor perturbations, $h_{ij}$, the interaction term reads~\cite{Parker:2009uva}
	\begin{equation}
		H_{I}=(2\mu)^{-1}h_{ij}\partial_{i}\partial_{j}\,.
	\end{equation}
	Consider the case that the cloud is dominated by the saturated state $|\psi_{1}\rangle$.  The interaction term induces the mixing of $|\psi_{1}\rangle$ and another state $|\psi_{2}\rangle$. 
	Since the gravitons are spin-2 particles, the mixing $\langle \psi_{1}|H_{I}|
	\psi_{2}\rangle$ vanishes if the difference of the magnetic quantum numbers $|m_{1}-m_{2}|\neq 2$. The wave function of the cloud is the linear summation of the two states $|\psi\rangle=c_{1}|\psi_{1}\rangle+c_{2}|\psi_{2}\rangle$, with the normalization condition $|c_{1}|^{2}+|c_{2}|^{2}=1$.
	The nearby GW sources, such as inspiral systems of compact binaries, provide the origin of external GWs. Since the distance of the GW sources is in general much larger than the radius of the cloud, external GWs can be treated as plane waves.
	Consider external GWs with circular polarization
	\begin{equation}
		h_{ij}(t,\mathbf{r})=h_{E}e^{i(\mathbf{k}\cdot\mathbf{r}-kt)}\mathrm{e}^{A}_{ij}+c.c.\,,
	\end{equation}
	where $h_{E}$ is the amplitude of external GWs, $A=R$,$L$ represents the polarization tensors of right-handed and left-handed GWs with $\varepsilon_{i j k} k_i \mathrm{e}_{k l}^A=i k \lambda_A \mathrm{e}_{j l}^A$, $\lambda_{R}=1$, $\lambda_{L}=-1$, $\varepsilon_{ijk}$ is the Levi-Civita tensor. The direction of GWs is assumed to be parallel to the BH angular momentum.
	In the interaction picture, 
	\begin{equation}\label{eq:schr}
		i\frac{d}{dt}\bigg(\begin{array}{c}	c_1(t)\\c_2(t)\end{array}\bigg)=\left(\begin{array}{c}	0\\ \eta e^{-i(\Delta E-\lambda_{A} k)t}\end{array}\begin{array}{c}	\eta e^{i(\Delta E-\lambda_{A} k)t}\\i\Gamma\end{array}\right)\left(\begin{array}{c}	c_1(t)\\c_2(t)\end{array}\right)\,,
	\end{equation}
	where $\Delta E$ is the energy difference of the two levels, the sign before $\lambda_{A}$ depends on the direction of GWs, $|\psi_{2}\rangle$ is an unstable state so that $\Gamma<0$.
	In the limit $|\Gamma c_{2}|\ll |\eta c_{1}|$, the instability of $|\psi_{2}\rangle$ hardly affect the solution of Eq.~\eqref{eq:schr} so we firstly neglect the $i\Gamma$ term and discuss this effect later. Since initially $|\psi_{1}\rangle$ is the dominant state, the initial condition is set to be $c_{1}=1$ and $c_{2}=0$. The parameter $\eta= \epsilon\alpha^{2}\mu h_{T}/2$ where $h_{T}=h_{E}+h_{S}$ is the total amplitude of GWs, the summation of both external GWs and GWs released from stimulated transitions. And the constant $\epsilon=0.07$ for $|\psi_{1}\rangle=|322\rangle$ and $|\psi_{2}\rangle=|100\rangle$, $\epsilon=0.14$ for $|\psi_{1}\rangle=|211\rangle$ and $|\psi_{2}\rangle=|21-1\rangle$. The transition rate is very low at the beginning so that $h_{S}$ is negligible compared to $h_{E}$, thereby $\eta$ is almost a constant. Then, Eq.~\eqref{eq:schr} has the same form as that obtained in Refs.~\cite{Baumann:2018vus,Berti:2019wnn} which considers the impact of a companion. Notably, right-handed GWs with the wavenumber $k=\Delta E$ result in resonant level transitions that produce identical gravitons as external GWs. The solution of Eq.~\eqref{eq:schr} becomes
	\begin{equation}\label{eq:solu}
		c_{1}(t)=\cos(\eta_{E} t)\,,\quad
		c_{2}(t)=i\sin(\eta_{E} t)\,,
	\end{equation}
	where $\eta_{E}\equiv \epsilon \alpha^{2}\mu h_{E}$. 
	Since the total occupation number of ultralight particles is conserved, the increase of $c_{1}$ and the decrease of $c_{2}$ imply the GW stimulated transitions from $|\psi_{1}\rangle$ to $|\psi_{2}\rangle$. The transitions between the level release energy into stimulated GWs, with the GW radiation power $\Delta E N_{0}\partial_{t}|c_{2}|^{2}$, where $N_{0}=M_{c}/\mu$ is the occupation number of the state and $M_{c}\equiv \sigma M$ is the total mass of the cloud.
	Since the wave functions are approximately restricted within the spherical region with the radius $r_{c}\approx 2n^{2}/(\alpha\mu)$, the GW radiation power equals the energy of GWs that escape that sphere per unit time, which is estimated by $r_{c}^{2} k^{2}h_{S}^{2}/2$. This implies the estimation of the amplitude of stimulated GWs 
	\begin{equation}\label{eq:hs}
		h_{S}=\left(\frac{2\Delta E N_{0}}{ r_{c}^{2}k^{2}}\partial_{t}|c_{2}|^{2}\right)^{1/2}\equiv \left(\kappa\partial_{t}|c_{2}|^{2}\right)^{1/2}\,.
	\end{equation}  
	
	\begin{figure}
		\includegraphics[width=3.2in]{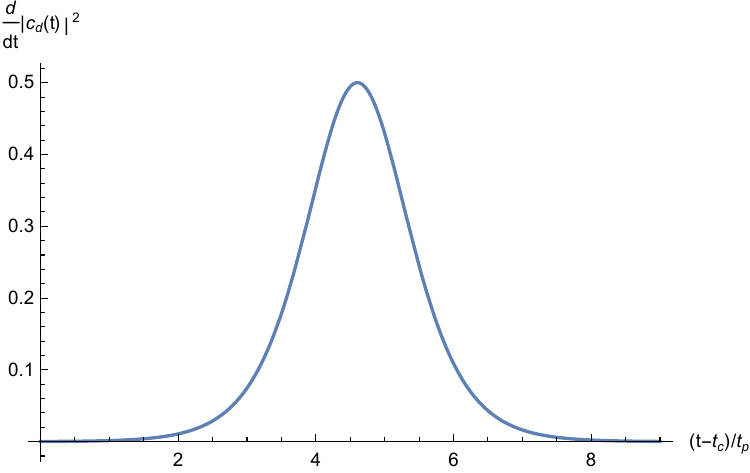}
		\caption{The evolution of the transition rate, which is proportional to the power of GW radiation. Here we adopt $\delta=0.01$ for illustration. This figure implies the timescale of GW radiation is comparable to $t_{p}$.}
		\label{fig:dtcd}
	\end{figure}
	
	As the transition rate increases with time, $h_{S}$ eventually becomes dominant so that $h_{T}\approx h_{S}$. In the cases that external GWs cease or the frequency shifts out of the resonance band, $h_{S}$ can also become dominant. Then, the solution of Eq.~\eqref{eq:schr} reads 
	\begin{equation}\label{eq:solu2}
		c_{2}(t)=i\frac{e^{(t-t_{c})/t_{p}}}{\sqrt{\delta^{-2}-1+e^{2(t-t_{c})/t_{p}}}}\,,\quad t_{p}=\kappa^{-1}\left(\frac{\eta}{h_{T}}\right)^{-2}\,,
	\end{equation}
	where $t_{c}$ denotes the beginning that $h_{S}$ becomes dominant and $\delta\equiv c_{2}(t_{c})$. To derive Eq.~\eqref{eq:solu2} we have used the normalizing condition $|c_{2}|^{2}=1-|c_{1}|^{2}$. The evolution of $c_{2}$ gradually changes from Eq.~\eqref{eq:solu} to Eq.~\eqref{eq:solu2} at around $t_{c}$. Eqs.~(\ref{eq:hs},\ref{eq:solu2}) together imply that $t_{c}\sim t_{p}$ and so that $\delta=\eta_{E}t_{p}$. Under the realistic conditions, the duration of external GWs may be smaller than $t_{p}$, i.e., $\delta<\eta_{E}t_{p}$. In the limit $c_{2}\ll 1$, $c_{2}(t)$ is proportional to $e^{t/t_{p}}$. The evolution of $\partial_{t}|c_{2}|^{2}$ is depicted in Fig.~\ref{fig:dtcd}, which is proportional to the GW radiation power. Fig.~\ref{fig:dtcd} indicates that most of the energy released into GWs within the period comparible to $t_{p}$. 
	Intuitively speaking, external GWs play the role of a kindling that ignites the system. Although the BH does not absorb the particles in $|\psi_{1}\rangle$ state, the cloud is an unstable system because of the self-generation of GWs.
	Eq.~\eqref{eq:solu2} also implies the total time from the initiation of external GWs to the termination of stimulated transitions
	\begin{equation}
		T\approx -\ln(\eta_{E}t_{p})t_{p}\,.
	\end{equation} 
	Because of the logarithmic dependence, $T$ is not sensitive to the duration and the strength of external GWs.
	
	The existence of external GWs is guaranteed by the widespread presence of compact binaries, as observed by the LIGO-Virgo collaboration. The GW frequency gradually increases during the inspiral phase due to GW production, covering a broad frequency range that includes the resonance band.
	Since $T$ only logarithmically depends on $h_{E}$, the inspiral systems with a large distance also have the chance to ignite the system, providing abundant sources of external GWs in the resonance band. The resonant frequency is in general orders of magnitude smaller than the GW frequency from mergers, in this case, allowing GWs from inspirals to be effectively treated as monochromatic. 
	
	
	Yet, we have neglected the instability of $|\psi_{2}\rangle$. If $|\psi_{2}\rangle$ is a strongly absorptive state such as $l_{2}=0$, the condition $|\eta c_{g}|\gg|\Gamma c_{2}|$ is vialated before $h_{S}>h_{E}$. The effects from level transitions and the absorption counteract each other. In this case, $c_{2}$ approaches a constant and $c_{g}$ decreases linearly with time, which means external GWs cannot ignite the system. This is the intrinsic property of $|\psi_{1}\rangle$ and $|\psi_{2}\rangle$, independent of the external GW strength. To successfully ignite the system, the condition $|\eta c_{g}|\gg|\Gamma c_{d}|$ must be satisfied before $h_{S}$ becomes dominant, which yields the threshold condition $\Gamma t_{p}\ll 1$.  
	
	Given that the majority of the energy is released within the timescale of $t_{p}$, the prediction of stimulated GWs is more similar to the GW events resulting from compact binary mergers. This signature is different from the typical continuous GW signals produced by boson annihilations.
	The energy difference between the levels is relatively small compared to the total energy of the cloud. The BH absorbs the particles in the unstable state $|\psi_{2}\rangle$. As a result, there is a suppression in both the total energy and the frequency of stimulated GWs.
	the instantaneous radiation power of stimulated GWs greatly surpasses that of boson annihilations, thereby facilitating the identification of individual sources.
	The estimation of the GW strength at the observer requires the divergence angle $\Delta\Omega$, which represents the difference in directions between stimulated GWs and external GWs.
	\begin{equation}\label{eq:h}
		h=\left(\frac{\Delta ED^{2}t_{p}}{2 N_{0}}\frac{\Delta\Omega}{4\pi}\right)^{-1/2}\,.
	\end{equation}
	The accurate result of $\Delta\Omega$ is beyond the scope of this work. However, Eq.~\eqref{eq:h} provides a reference value of $h$ as long as $\Delta\Omega$ is not too small. The event rate and $h$ both depend on $\Delta\Omega$, whereas the total energy of stimulated GWs is irrelevant to $\Delta\Omega$. 
	
	
	
	Following the completion of the transitions, stimulated GWs carry away the energy of the system, causing a sudden decrease in $h_{T}$. Consequently, the transitions stimulated by GWs becomes irreversible, distinguishing it from the that triggered by a companion. The rapid transition process results in a high occupation number in an unstable state. Ad the BH absorbs the particles in $|\psi_{2}\rangle$, the amplitude of GWs from boson annihilations also decreases exponentially. This phenomenon serves as additional evidence for the existence of stimulated GWs, complementing direct observations. For $m_{2}<0$, the angular momentum of $|\psi_{2}\rangle$ is opposite to that of the BH. Then, the spin of the BH may be much lower than the threshold given in Ref.~\cite{Arvanitaki:2010sy} when the absorption completes.
	
	
	\emph{Examples and observational applications}.
	We calculate the reference values of the observables in terms of the basic parameters
	\begin{equation}
		\begin{aligned}
			h&\approx 6\times10^{-26+A}\frac{\epsilon}{0.1}\frac{\sigma}{0.01}\left(\frac{n_{1}}{2}\right)^{-2}\left(\frac{\alpha}{0.1}\right)^{3-A}\left(\frac{D}{\mathrm{Mpc}}\right)^{-1}\left(\frac{\Delta\Omega}{4\pi}\right)^{-\frac{1}{2}}\frac{M}{10M_{\odot}}\,,\\
			f&\approx3\times10^{2-A}\left(\frac{\alpha}{0.1}\right)^{A+1}\left(\frac{M}{10M_{\odot}}\right)^{-1}\,\mathrm{s}^{-1}\,,\\
			t_{p}&\approx 6\times10^{9-A}\left(\frac{\epsilon}{0.1}\right)^{-2}\left(\frac{\sigma}{0.01}\right)^{-1} \left(\frac{n_{1}}{2}\right)^{4}\left(\frac{\alpha}{0.1}\right)^{A-8}\frac{M}{10M_{\odot}}\,\mathrm{s}\,,\\
			\frac{T}{t_{p}}&\approx 28+2.3A+\ln\left(\frac{\epsilon}{0.1}\frac{\sigma}{0.01}\left(\frac{n_{1}}{2}\right)^{-4}\left(\frac{\alpha}{0.1}\right)^{5-A}\left(\frac{h_{E}}{10^{-20}}\right)^{-1}\right)\,,
		\end{aligned}
	\end{equation}
	where $D$ is the distance of the boson cloud. We have applied the estimation $\Delta E\approx \alpha^{A}\mu$ by neglecting the constant coefficients, where $A=2$ for Bohr splitting, $A=4$ for fine splitting and $A=5$ for hyperfine splitting. The population of individual  Kerr BHs is much higher than the formation rate of compact binaries, which leads to a considerable event rate, offering the possibility of detecting these sources even at relatively short distances.
	
	The frequency of GWs from level transitions, which is suppressed by the factor $\alpha^{A}$, is orders of magnitude lower than that from boson annihilations. The observation of stimulated GWs expands the mass range of ultralight bosons that can be detected by each GW observer. For instance, the sensitivity band of the LIGO-Virgo collaboration~\cite{TheLIGOScientific:2014jea} spans from $10$ Hz to $10^{3}$ Hz, corresponding to GW frequencies from ultralight field annihilation with a mass range of $10^{-13}$ eV to $10^{-12}$ eV~\cite{Yuan:2022bem}. The absence of a stochastic GW background imposes constraints on ultralight fields within this mass range. The observation of stimulated GWs from transitions provides the possibility for ground-based laser interferometers to constrain ultralight fields with masses around $10^{-11}$ eV, $10^{-9}$ eV, and $10^{-8}$ eV for Bohr, fine, and hyperfine transitions with $\alpha=0.1$. (The superradiance of boson fields with masses $\gtrsim 10^{-11}$ eV requires primordial BHs, which have been proposed as important candidates for dark matter. Primordial BHs formed in the early Universe exhibit a broad mass spectrum, enabling the detection of light boson fields across an extensive range of masses. This mechanism predicts the presence of a background of stimulated GWs that originated prior to the formation of stars, exhibiting large redshift.) Stimulated GWs from clouds of supermassive BHs also offer a potential explanation for the nHz signal recently reported in pulsar timing array experiments~\cite{Antoniadis:2023ott,Reardon:2023gzh,Xu:2023wog,NANOGrav:2023hvm}.
	
	The impact of GWs significantly alters the distribution function of the clouds, consequently affecting the prediction of the stochastic GW background resulting from boson annihilations. As previously discussed, compact binaries are the most likely source of external GWs. The observation of stimulated GWs predicts an additional GW signal arising from the merger events of these binaries, which contribute to the external GWs.
	
	In the case $E_{2}>E_{1}$, particles in the $|\psi_{1}\rangle$ state have the ability to absorb left-handed gravitons with $k=\Delta E$ and transition to the $|\psi_{2}\rangle$ state. However, this absorption process leads to a reduction in the strength of external GWs as well as the transition rate, resulting in a prolonged timescale for the absorption process. This effect can potentially be observed through the decrease in the amplitude of continuous GWs originating from sources such as the inspiral of compact binaries and neutron stars. Similar to the ionization of a hydrogen atom, boson particles are also capable of absorbing the energy of gravitons with $k>\mu-E_{nlm}$ and subsequently becoming free particles. See also Refs.~\cite{Baumann:2021fkf,Baumann:2022pkl} for the ionization induced by a companion of the BH.
	
	According to Ref.~\cite{Tong:2022bbl}, a companion with an angular velocity outside the resonance band has the ability to terminate superradiance. We can draw similar conclusions in our case as well. It has been found that external GWs with $k$ significantly higher or lower than $\Delta E$ can also lead to the termination of superradiance. For instance, in the vicinity of supermassive BH binaries, strong tensor perturbations can be considered as a stationary background. The termination of the superradiance of the $|322\rangle$ state requires $\eta>2\times 10^{-7}\, \mathrm{s}^{-1}$, or equivalently, for $h_{E}>10^{-9}$ when $\alpha=0.1$ and $M=100M_{\odot}$.
	
	GWs emitted during the mergers of compact binaries carry away linear momentum, resulting in a recoil velocity of the remnant BH~\cite{Gonzalez:2006md}. Similarly, the stimulated transitions in gravitational atoms also induce a recoil velocity of the BH due to the production of directed GWs. The total energy of GWs, estimated as $\sigma\alpha^{A}M$, implies that the maximum recoil velocity can reach $10^{-3}$ times the speed of light for Bohr splitting.

	\emph{Conclusion and discussion}.
	In this work, we investigate the stimulated GW radiation in gravitational atoms. We consider BHs surrounded by boson clouds formed through superradiance. We discover that external GWs with an angular frequency close to the energy difference between two levels can induce resonant transitions. These transitions, in turn, generate GWs of the same frequency and direction. Remarkably, the stimulated GWs exponentially amplify the transition rate, facilitating their completion within a much shorter time compared to the cloud's lifetime. This mechanism predicts a new class of GWs that is anticipated to be detected by various GW observatories. The observation of such GWs would expand the mass range of ultralight bosons accessible to GW observations. The stimulated transitions can also be indirectly verified by observing the continuous GW signal from boson annihilations, which exhibits an exponentially decaying amplitude. Additionally, the stimulated absorption of GWs could be detected by observing the reduction in the amplitude of continuous GWs. Since stimulated level transitions significantly shorten the lifetime of the clouds, the distribution of the clouds and the GW signal from boson annihilations require reevaluation.
	
	In addition to scalar fields, vector~\cite{Konoplya:2005hr,Rosa:2011my,Frolov:2018ezx,Dolan:2018dqv} and tensor fields~\cite{Babichev:2013una,Brito:2013yxa,Brito:2020lup} can also form clouds via superradiance. Furthermore, superradiant instabilities can be induced by various astrophysical objects, including stars~\cite{Richartz:2013unq,Cardoso:2015zqa}, neutron stars~\cite{Day:2019bbh}, boson stars~\cite{Gao:2023gof}, and BH binaries~\cite{Wong:2019kru,Ribeiro:2022ohq}. These systems may also give rise to stimulated GW radiation, revealing a broader range of phenomena that we leave for future exploration.
	
	
	\emph{Acknowledgments}
	We sincerely thank Zong-Kuan Guo, Jun Zhang, Shao-Jiang Wang, Chang Liu and Qiang Jia for the fruitful discussions.
	This work is supported in part by the National Key Research and Development Program of China Grants No. 2020YFC2201501 and No. 2021YFC2203002, in part by the National Natural Science Foundation of China Grants No. 12105060, No. 12147103, No. 12235019, No. 12075297 and No. 12147103, in part by the Science Research Grants from the China Manned Space Project with NO. CMS-CSST-2021-B01,
	in part by the Fundamental Research Funds for the Central Universities. 
	
	\section{Supplemental material: the Superradiance and Boson Cloud}
	
	Consider the system of the ultralight scalar field $\Psi$ with mass $\mu$ surrounding a rotating BH with mass $M$ and dimensionless spin parameter $a$. The Klein-Gordon equation of $\Psi$ reads
	\begin{equation}
		\left(g^{\alpha\beta}\partial_{\alpha}\partial_{\beta}-\mu^{2}\right)\Psi(t, \mathbf{r})=0\,,
	\end{equation}
	where $g^{\alpha\beta}$ stands for the
	contravariant Kerr metric. In the non-relativistic limit $\alpha\equiv M\mu\ll 1$, $\Psi(t, \mathbf{r})$ can be conveniently decomposed into
	\begin{equation}
		\Psi(t, \mathbf{r})=\frac{1}{\sqrt{2 \mu}}\left[\psi(t, \mathbf{r}) e^{-i \mu t}+\psi^*(t, \mathbf{r}) e^{+i \mu t}\right]\,,
	\end{equation}
	where the evolution timescale of $\psi$ is much larger than $\mu$.
	Neglecting the higher order terms of $r^{-1}$, the equation of $\psi(t, \mathbf{r})$ has the same form as the Schrödinger equation of the hydrogen atom
	\begin{equation}
		i \frac{d}{dt} \psi(t, \mathbf{r})=\left(-\frac{1}{2\mu} \nabla^2-\frac{\alpha}{r}\right) \psi(t, \mathbf{r})\,,
	\end{equation}
	where $\alpha$ also represents the fine-structure constant of the gravitational atom.
	In the limit $r\gg M$, the large-distance approximation of the eigenstates reads 
	\begin{equation}
		\psi_{n l m}(t,\boldsymbol{r})=e^{-i\left(\omega_{n l m}-\mu\right) t}R_{n l}(r) Y_{l m}(\theta, \phi) \,,
	\end{equation}
	where $Y_{l m}(\theta, \phi)$ is the spherical harmonics and
	\begin{equation}
		R_{nl}(r)=\sqrt{\left(\frac{2}{n}\right)^{3}\frac{(n-l-1)!}{2n(n+l)!}}e^{-r/n}\left(\frac{2r}{n}\right)^{l}L_{n-l-1}^{2l+1}(2r/n)\,,
	\end{equation}
	where $L$ denotes the generalized Laguerre polynomial. Due to the absorptive horizon and dissipative ergoregion of the BH, the eigenfrequencies are complex numbers
	\begin{equation}
		\omega_{nlm}=E_{nlm}+i\Gamma_{nlm}\,.
	\end{equation}
	The energy levels read~(up to the hyperfine splitting)~\cite{Baumann:2019eav}
	\begin{equation}
		\begin{aligned}
			E_{nlm}=\mu\left(1-\frac{\alpha^{2}}{2n^{2}}-\frac{\alpha^{4}}{8n^{4}}+\frac{(2l-3n+1)\alpha^{4}}{n^{4}(l+\frac{1}{2})}+\right.\\
			\left.\frac{2am\alpha^{5}}{n^{3}l(l+\frac{1}{2})(l+1)}+\mathcal{O}(\alpha^{6})\right)\,,
		\end{aligned}
	\end{equation}
	which indicates the energy splitting is proportional to $\alpha^{2}$ for Bohr splitting, $\alpha^{4}$ for fine splitting and $\alpha^{5}$ for hyperfine splitting.
	In the limit $\alpha\ll 1$, $\Gamma_{nlm}$ has an analytical expression given by Detweiler~\cite{Detweiler:1980uk}
	\begin{equation}\label{eq:gamma}
		\Gamma_{n l m}=\frac{2 r_{+}}{M} C_{n l m}(a, \alpha)\left(m \Omega_H-\omega_{nlm}\right) \alpha^{4 l+5}\,,
	\end{equation}
	where $r_{+}=M+\sqrt{M^2-a^2}$ is the
	Boyer-Lindquist horizon radius, and $\Omega_{H}=\frac{a}{2Mr_{+}}$
	is the angular velocity of the outer horizon, and the constant
	\begin{equation}
		\begin{gathered}
			C_{n l m}(a, \alpha)=\frac{2^{4 l+1}(n+l) !}{n^{2 l+4}(n-l-1) !}\left[\frac{l !}{(2 l) !(2 l+1) !}\right]^2 \times \\
			\prod\left[j^2\left(1-a^2\right)+\left(a m-2 \tilde{r}_{+} \alpha\right)^2\right] \,.
		\end{gathered}
	\end{equation}
	The increase of the occupation number of the states with $\Gamma_{nlm}$ results in the descent of the BH angular momentum. The superradiance ceases when the condition $m \Omega_H>\omega_{nlm}$ is violated.
	
	\bibliography{laser}
\end{document}